\documentstyle[sprocl,epsfig]{article}

\def\PRD{{\em Phys. Rev.} D}

\newcommand{\newc}{\newcommand}
\newc{\hc}{{\it {h.c.}}}
\newc{\ie}{{\it  {i.e.}}}
\newc{\eg}{{\it  {e.g.}}}
\newc\etal{{\it {et al.}}}
\newc{\etc}{{\it {etc.}}}

\newcommand\lsim{\mathrel{\rlap{\lower4pt\hbox{\hskip1pt$\sim$}}
    \raise1pt\hbox{$<$}}}
\newcommand\gsim{\mathrel{\rlap{\lower4pt\hbox{\hskip1pt$\sim$}}
    \raise1pt\hbox{$>$}}}

\newc{\mx}{M_{GUT}}
\newc{\gx}{g_{GUT}}     \newc{\alphax}{\alpha_{\rm GUT}}

\newcommand\mchi{m_{\chi}}

\newc{\tanb}{\tan\beta}
\newc{\mtop}{m_t}
\newc{\mbot}{m_b}
\newc{\mtau}{m_{\tau}}
\newc{\htop}{h_t}
\newc{\hbot}{h_b}
\newc{\htau}{h_{\tau}}
\newc{\stopq}{{\widetilde t}}   
\newc{\mstop}{m_{\stopq}}  
\newc{\stp}{\stopq}
\newc{\stopl}{\widetilde t_L}
\newc{\stopr}{\widetilde t_R}
\newc{\stopone}{\widetilde t_1}
\newc{\stoptwo}{\widetilde t_2}

\newc{\mstopq}{m_{\tilde t}}
\newc{\mstopl}{m_{\tilde t_L}}
\newc{\mstopr}{m_{\tilde t_R}}
\newc{\mstopone}{m_{\tilde t_1}}
\newc{\mstoptwo}{m_{\tilde t_2}}
\newc{\stau}{{\widetilde \tau}}
\newc{\staul}{\widetilde \tau_L}
\newc{\staur}{\widetilde \tau_R}
\newc{\stauone}{\widetilde \tau_1}
\newc{\stautwo}{\widetilde \tau_2}

\newc{\azero}{A_0}
\newc{\bzero}{B_0}
\newc{\muzero}{\mu_0}
\newc{\sgnmu}{{\rm sgn}\,\mu}

\newc{\cachigamma}{C_{a\chi\gamma}}

\newc{\caww}{C_{aWW}}                   \newc{\cayy}{C_{aYY}}
\newc{\sthw}{\sin\theta_W}              \newc{\cthw}{\cos\theta_W}
\newc{\bino}{\widetilde B}              \newc{\wino}{\widetilde W_3}
\newc{\higgsinob}{{\widetilde H}^0_b}   \newc{\higgsinot}{{\widetilde H}^0_t}
\newc{\mcharone}{m_{\charone}}  \newc{\charone}{\chi_1^\pm}
\newc{\sfermion}{\widetilde f}
\newc{\msf}{m_{\sfermion}}

\newc{\abund}{\Omega h^2}
\newc{\omegachi}{\Omega_\chi}
\newc{\abundchi}{\Omega_\chi h^2}
\newc{\rhocrit}{\rho_{crit}}
\newc{\rhochi}{\rho_{\chi}}

\newc{\sigmaann}{\sigma_{\rm ann}}

\newc{\sigmap}{\sigma_p}

\newc{\mwimp}{m_{\rm WIMP}}     \newc{\rhowimp}{\rho_{\rm WIMP}}

\newc{\mplanck}{M_{\rm P}}              \newc{\mgut}{M_{\rm GUT}}
\newc{\mz}{m_{Z}}                       \newc{\mw}{m_{W}}

\newc{\xf}{x_f}
\newc{\jxf}{J({\xf})}
\newc{\VEV}[1]{\langle #1 \rangle}

\newcommand\gev{\,\mbox{GeV}}

\newcommand\kev{\,\mbox{keV}}

\newc{\ra}{\rightarrow}
\newc{\beq}{\begin{equation}}
\newc{\eeq}{\end{equation}}

\newc{\ekd}{\; {\rm event}/kg/{\rm day}}
\newc{\ekds}{\; {\rm events}/kg/{\rm day}}
\newc{\ekkds}{\; {\rm events}/kg/keV/{\rm day}}
\newc{\kgday}{\;\mbox{kg$\times$day}}
\newc{\pb}{\,\mbox{pb}}
\newc{\rhozerothree}{\rho_{0.3}}
\newc{\gevcmcube}{\,\mbox{GeV/cm$^3$}}

\newc{\sigchin}{\sigma(\chi N)}

\def\PRD#1#2#3{Phys. Rev. D {\bf#1} (19#2) #3}

\begin{document}

\begin{titlepage}
\pagestyle{empty}
\baselineskip=21pt
\rightline{CERN--TH/2001-060}
\vskip 0.5in
\begin{center}
{\large {\bf Hide and Seek with Neutralino Dark Matter 
WIMP\footnote{Invited
talk at the 3rd International Conference on Particle Physics and the
Early Universe (COSMO-99), Trieste, Italy, 27 September - 3 October,
1999.}
}}
\end{center}
\begin{center}
\vskip 0.05in
{\bf Leszek Roszkowski}$^{1,2}$\\

\vskip 0.05in
{\it
$^1${Department of Physics, Lancaster University, Lancaster LA1
4YB, England}\\
$^2${TH Division, CERN, CH-1211 Geneva 23, Switzerland}\\
}
\vskip 0.5in
{\bf Abstract}
\end{center}
\baselineskip=18pt 
\noindent 
As experimental sensitivity increases, one is
approaching the range of WIMP-nucleon interaction strengths
characteristic of neutralinos. But this continuing progress also unearths
new experimental challenges and uncertainties.
\vfill
\vskip 0.15in
\leftline{CERN--TH/2001-060}
\leftline{February 2001}
\end{titlepage}

\vspace*{1.5cm}

\title{Hide and Seek with Neutralino Dark Matter 
WIMP
\protect\,\footnote{Invited
talk at the 3rd International Conference on Particle Physics and the
Early Universe (COSMO-99), Trieste, Italy, 27 September - 3 October,
1999.}
}

\author{Leszek Roszkowski}

\address{Department of Physics, Lancaster University, 
Lancaster LA1 4YB, England\\E-mail: l.roszkowski@lancaster.ac.uk}


\maketitle
\abstracts{
As experimental sensitivity increases, one is
approaching the range of WIMP-nucleon interaction strengths
characteristic of neutralinos. But this continuing progress also unearths
new experimental challenges and uncertainties.}

\section{Introduction}

The search for dark matter (DM) in the Universe is now in full swing.
The underlying assumption is that the Milky Way is immersed in an
extended, approximately spherical halo of WIMPs. The local (\ie, at
our Sun's distance from the Galactic center) halo density is estimated
at $0.3\gev/{\rm cm^3}$ with a factor of two or three
uncertainty.\cite{jkg} This translates to about 3000 WIMPs with mass
$\mchi=100\gev$ per cubic meter. With typical velocities in the range
of a few hundred $km/s$, the resulting flux of WIMPs is actually quite
large, $\Phi=v\rho_\chi/\mchi\approx10^{9}\left(100\gev/\mchi\right)
{\rm \chi\,s}/m^2/sec$.

In the case of very-well motivated neutralino WIMPs of supersymmetry, 
their cross section $\sigma(\chi N)$ for elastic
scattering from a target nucleus $N$ is expected to be typically very
small, roughly below $10^{-6}\pb$. This is because the elastic cross
section is related by crossing symmetry to the cross section
$\sigmaann$ of neutralino annihilation in the early Universe 
which has to be only of a fraction of weak
interaction strength in order to give $\abundchi\sim1$.

Such small cross sections are obviously an enormous challenge to
experimentalists aiming to search for dark matter. Several strategies
have been developed to boost one's chances for detecting the elusive
WIMP. However, as is often the case with novel experiments exploring
uncharted cross section ranges, a number of new questions arise and
need to be resolved.

In this talk I will address some of the issues surrounding WIMP
searches. In particular, I will focus on a recent claim of the DAMA
Collaboration of a possible WIMP signal in their data and on a new
limit from the CDMS. I will also compare this with updated predictions
from supersymmetry (SUSY). First some basics.

\section{Basics}
Measurements of the last few years have put significant constraints on
the allowed range of the WIMP relic abundance $\abundchi$.  Current
estimates of the lower bound on the age of the Universe lead to
$\abund\lsim0.25$.  Recent results from high-redshift supernovae type
Ia imply $\Omega_{\rm matter}\simeq0.3$. The Hubble parameter is now
constrained to $0.65\pm0.1$. Since $\Omega_{\rm baryon}
h^2\lsim0.015$, one obtains $0.1\lsim \Omega_{\chi} h^2\lsim0.15$ or
so. I will use this range as a favored one. Some authors allow for
somewhat larger values of $\abundchi$ up to 0.3 or 0.4. This will not
affect my basic conclusions. On the other hand, values of $\abundchi$
below 0.1 or so, while not excluded, are questionable. They would
imply that besides WIMPs there exists yet another DM component of
matter.  At the very least, requiring that galactic halos are made
mostly of WIMPs leads to $\abundchi\gsim0.025$.

Neutralino WIMP elastic scattering cross section off a nucleus
receives contributions from effective scalar (or coherent) and spin
(or incoherent) interaction terms. For targets with sufficiently large
mass number $A\gsim30$ the former dominate so I will only concentrate
on them below.

The coherent part of the cross section is described by an effective
scalar coupling between the WIMP and the nucleus is proportional to
the number of nucleons in the nucleus. It receives dominant tree-level
contribution from scattering off constituent quarks, $\chi q\ra \chi
q$ mainly via Higgs and squark exchange.  The resulting cross section
for scalar neutralino-nucleus interactions is $\sigma^{scalar}(\chi N)
\sim G_F^2 {{\mchi^2 m_N^2}/{(m_N +\mchi)^2}} A^2$.
It is often 
convenient to express $\sigma^{scalar}(\chi N)$ in terms of the
WIMP-proton cross section $\sigmap$: $\sigma^{scalar}(\chi N) = A^2
({\mu_A^2}/ {\mu_p^2}) \sigmap$
where $\mu_i= \mchi m_i/(\mchi+m_i)$ is the reduced mass. This allows one
to compare
limits derived by different experiments which use different target
materials. Second, theoretical calculations in specific (\eg, SUSY)
models give predictions for $\sigmap$ which can be
next directly compared with experimental results.

\subsection{Recent Experimental Results
} \label{seasonal:sec}
\paragraph{DAMA and Annual Modulation.}
One interesting strategy for detecting a WIMP signal is to look for
yearly time variation in the measured energy spectrum. 
It has been pointed out \cite{dfs86,ffg88} that a halo WIMP signal
should show a periodic effect due to the Sun's motion through the
Galactic halo, combined with the Earth's rotation around the Sun. The
peaks of the effect are on the 2nd of June and half a year later.

The effect, called ``annual modulation'', would provide a convincing
halo WIMP signal. Unfortunately, in SUSY models the effect is usually
small, $\lsim5\%$.\cite{jkg,bb97}  With the absolute event rates
being already very small, it is going to be a great challenge to
detect it.

Now I would like to make some comments about possible evidence for a
WIMP signal in annual modulation that has been claimed by the DAMA
Collaboration.  Based on the combined statistics of $57,986\kgday$ of
data collected in a NaI detector since November '96, the Collaboration
has reported~\cite{dama2000} a statistically significant
($4\,\sigma$~CL) effect which could be caused by an annual modulation
signal which, according to DAMA, corresponds to
\begin{eqnarray}
\mchi &=& 52\gev^{+10}_{-8}\gev,\\
\xi_{0.3}\sigmap &=& 
7.2^{+0.4}_{-0.9}\times10^{-6}\pb,
\label{damarange:eq}
\end{eqnarray}
where $\xi_{0.3}=\rhochi/\rhozerothree$ stands for
the local WIMP mass density $\rhochi$ normalized to
$\rhozerothree=0.3\gevcmcube$. (See Fig.~1
and also Fig.~4 in
Ref.~5 
for a $3\sigma$ signal region in the
($\mchi,\xi_{0.3}\sigmap$) plane.) When the previously obtained upper
limit~\cite{dama96} is included in the analysis, DAMA obtains 
$\mchi = 44\gev^{+12}_{-9}\gev$ and $\xi_{0.3}\sigmap=
5.4\pm1.0\times10^{-6}\pb$ at $4\,\sigma$~CL. According to DAMA, 
the new analysis is consistent with and confirms
the Collaboration's earlier claim~\cite{dama98two} based on
$19,511\kgday$ of data for the presence of the
signal corresponding to $\mchi = 59\gev^{+17}_{-14}\gev$ and $\xi_{0.3}\sigmap=
7.0^{+0.4}_{-1.2}\times10^{-6}\pb$ at $99.6\%$~CL.

The claimed effect comes from a few lowest bins of the scintillation
light energy, just above the software threshold of $2~\kev$, and
predominantly from the first bin ($2 - 3~\kev$). This is indeed what
{\em in principle} one should expect from the annual modulation
effect.  DAMA appears confident about the presence of the effect in
their data, and claims to have ruled out other
possible explanations, like temperature effects, radon contamination
or nitrogen impurities. According to DAMA, the effect is caused by
single hit events
(characteristic of WIMPs unlike neutron or gamma background)
with proper modulation of about one year, peak around June, and small
enough amplitude of the time dependent part of the
signal. 

\begin{figure}[t!]      
\centering
\centerline{\epsfig{file=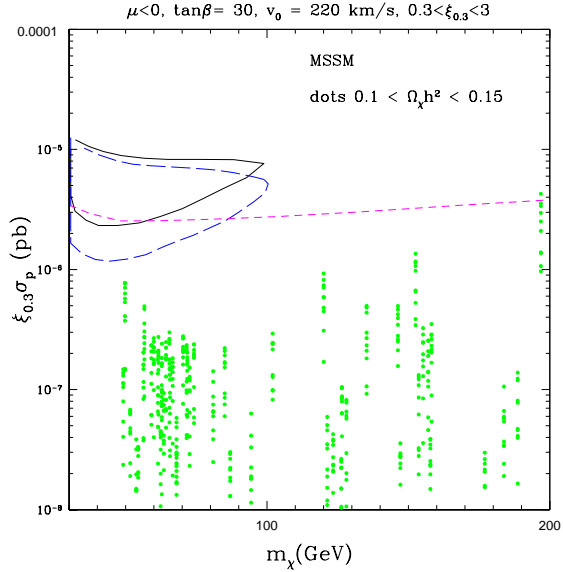,height=3in,width=3in}}
\bigskip
\caption{\small The allowed DAMA $3\sigma$ region without (solid) and with
(long-dash) DAMA's upper limit from 1996 imposed. The $90\%$~CL upper
limit from CDMS is shown as short dash. Thick dots correspond to a
scan of the minimal SUSY parameter space.\hfil }
\end{figure}

Nevertheless, several experimental questions remain and cast much
doubt on the validity of the claim. Here I will quote some of those
which I find particularly important to clarify. First, as stated
above, the claimed effect comes from the lowest one or two energy
bins. This is indeed what one should expect from an annual modulation
signal. But is the effect caused by just one or two energy bins
statistically significant? This is especially important in light of
the fact that the shape of the differential energy spectrum $dR/dE$ in
the crucial lowest energy bins as measured by DAMA is very different
from the one measured by Gerbier, \etal~\cite{gerbiernoteone} for
the same detector material (NaI). In Ref.~8 
the
corrected-for-efficiency $dR/dE$ is about $10\ekkds$ at 3\kev,
decreasing monotonically down to about $2\ekkds$ at 6\kev. (See
Fig.~15 in Ref.~8 
.)  In contrast, DAMA's spectrum
shows a dip down to $1\ekkds$ at $2\kev$, above which it increases to
nearly $2\ekkds$ at $4\kev$.\cite{dama2000} It is
absolutely essential for the controversy of the shape of $dR/dE$ in
the lowest energy bins to be resolved. Furthermore, DAMA's data from
the second run ($\sim15000\kgday$) shows that the background in the
crucial lowest bin ([2,3]\kev) is only about half or less of that in
the next bins.\cite{gaitskell:taup} One may wonder why this would be the case.
Examining more closely the data in the constituent nine NaI crystals,
one finds a rather big spread in the event rates.\cite{gaitskell:taup} In
detector~8, in the lowest energy bin one finds no contribution from the
background whatsoever! 

Two other groups which have also used NaI have
reported~\cite{gerbiernoteone,ukdmcanomevents} robust evidence of
events of unexpected characteristics and unknown origin. The data of
both teams has been analysed using a pulse shape analysis (PSA).
A small but statistically significant component
was found with the decay time even shorter than the one expected from
WIMPs or neutrons. While the population of those events appears to be
too small to explain DAMA's effect, a question remains not only about
their origin (contamination?, external effect?)  but also how they
contribute to the energy spectrum in the crucial lowest bins. DAMA
claims not to have seen such events.

It is also worth noting that in Ref.~11 
annual
modulation was reanalyzed for germanium and NaI. It was concluded that
the effect would be too small to be seen with current
sensitivity. Particularly illuminating is Fig.~6.a where DAMA's data
from Ref.~12 
(run~I) was re-plotted along with an
expected signal for the modulated part of the spectrum for the central
values of the ranges of the WIMP mass and cross section ($\sigmap$)
selected by DAMA. One can hardly see any correlation between the data
and the expected signal.

A controversy over DAMA's alleged signal is of experimental
nature. One may therefore hope that it will soon be definitively
resolved. It would be of particular importance for another experiment
using different detector material to put DAMA's claim to test.

\paragraph{CDMS.}
The CDMS cryogenic detection experiment using germanium and silicon
crystals at Stanford has recently published~\cite{cdms-2000} a new
limit on scalar WIMP-proton cross section. The new result is based on
the total of $10.6\kgday$ of data collected a current shallow site
(17~mwe) at Stanford during 1999. A powerful event-by-event
discrimination method allows CDMS to reach a sensitivity matching that of
DAMA with only less than $0.2\%$ of DAMA's statistics. 

The new $90\%$~CL CDMS limit is plotted as a solid line in
Fig.~1. It excludes most of the signal region
claimed by DAMA. In particular, it fully rules out the previous
$2\,\sigma$ region based on the combined data of $19,511\kgday$ from
runs~I and~II~\cite{dama98two} and the new $3\,\sigma$ region at more
than $84\%$~CL.

In the CDMS data most events are due to background electron and gamma
hits. These are rejected by a powerful discrimination technique which
allows a simultaneous measurement of ionization (background) and
phonons due to nuclear recoil (WIMPs and background neutrons). After
cuts (including a $10\kev$ recoil energy cut), 13 recoil events remained
which the Collaboration identified as neutrons. This corresponds to
about $1.2\ekd$ which would be only about $1/3$ of the number of WIMPs
needed to reproduce the DAMA annual modulation signal. In other words,
even in the highly unlikely case that all the 13 CDMS neutrons were
actually misidentified WIMPs, the two experiments would remain
incompatible. 

The CDMS experiment is currently collecting more data at Stanford and
during the next year will move to a deep site at Soudan. It is
expected that during the next few year this and other experiments (in
particular UKDMC-Xe, upgraded DAMA, GENINO and CRESST-II) are
likely to reach the sensitivity of $\sim10^{-8}\pb$ (for WIMP masses
up to a few hundred~$\gev$). In the perspective of the next decade the
range down to $\sim10^{-10}\pb$, or maybe even below, is hoped to be
probed. (For a comprehensive compilation, it is worth visiting a
particularly useful plotting facility of Gaitskell \&\ Mandic which is
now available on the Web: http://cdms.berkeley.edu/limitplots/.)

\subsection{Implications for SUSY and Theoretical Uncertainties}
It is interesting to compare the new experimental results with
predictions of minimal supersymmetry. For each choice of SUSY
parameters one can compute both the neutralino relic abundance and the
cross section $\sigmap$. A broad scan shows a number of points quite
close to the reach of current experimental sensitivity. This is
illustrated in Fig.~1.\cite{bgr99} Typically large
enough values of $\tan\beta$ are needed to increase $\sigmap$. SUSY
configurations corresponding to the cosmologically favored range
$0.1<\abundchi<0.15$ are denoted by thick points.

It is clear that current experiments have not quite yet reached the most
favored region of minimal SUSY neutralino if one assumes a current
cosmological range of a WIMP relic abundance. Relaxing the lower bound
down to truly very low values (0.025) increases corresponding cross
sections up and above current experimental sensitivity as shown in
Fig.~1.\cite{bgr99}

One should remember that there are other factors which may influence
both experimental results and theoretical predictions. One is the
shape and peak of the halo WIMP velocity distribution. The above
results have been obtained assuming a Maxwellian distribution peaked
around $v_0=220\, km/s$. For example, varying $v_0$ within a
reasonable range (say, $\pm50\, km/s$) leads to a significant
enlargement of the WIMP mass range selected in the DAMA run I+II
region~\cite{dama98two} as was first shown in Ref.~15. 
It has
also some, although relatively small effect, on the limits. It is also
plausible that the halo density distribution is not smooth. If so, the
WIMP density in the solar neighbourhood could be very different from
the average one.

Furthermore, quark mass inputs in calculating the scalar cross-section
for the neutralino-nucleus elastic scattering are fraught with some
errors, as mentioned above. The effect of the latter has been recently
re-analysed in Ref.~16 
and found not to affect the overall
scalar cross-section by more than a factor of a few at most.

It is clear that today's experiments are now only reaching the
sensitivity required to begin testing predictions coming from
SUSY. What I find promising is that several experiments using
different detector materials and often different methods of [attempts
at] distinguishing signal from background will explore a large
fraction of the SUSY parameter space within the next few years. 
Especially reassuring would be an observation of a positive signal in more
than type of DM detector, and/or for measuring also the directionality
of the signal, although many experimentalists would probably
remark that I am asking for too much. I remain an unrepentant optimist.

\section*{Acknowledgements}
I am greatly indebted to Goran Senjanovi{\' c} and Alexei Smirnov and
other members of the Local Organising Committee for arranging a
fruitful and inspiring meeting. My thanks also go to Rick Gaitskell
for illuminating comments about experimental subtleties surrounding
WIMP searches.

\end{document}